\documentclass[jphysa,amsmath]{iopart}

\usepackage{amsmathred}
\usepackage{color}
\usepackage{graphicx}
\usepackage{amsthm}
\usepackage{amsfonts}
\usepackage{amssymb}
\usepackage{braket}
\usepackage{times,txfonts}
\usepackage{subfigure}
\usepackage{hyperref}
\usepackage{url}           
\usepackage{bm}            
\usepackage{amsbsy}
\usepackage{dsfont}
\usepackage{braket}
\usepackage{cite}

\newtheorem*{thm*}{Theorem}

\theoremstyle{definition}

\newcommand{\EE}{\mathcal{E}}

\begin{document}

\title{Quantum coherence fluctuation relations}

\author{Benjamin Morris and Gerardo Adesso}
\address{{Centre for the Mathematics and Theoretical Physics of Quantum Non-Equilibrium Systems,\\ School of Mathematical Sciences, University of Nottingham, \\
University Park, Nottingham NG7 2RD, United Kingdom}\\ Email: \href{mailto:benjamin.morris@nottingham.ac.uk}{benjamin.morris@nottingham.ac.uk}}

\begin{abstract}
We investigate manipulations of pure quantum states under incoherent or strictly incoherent operations assisted by a coherence battery, that is, a storage device whose degree of coherence is allowed to fluctuate in the process. This leads to the derivation of fluctuation relations for quantum coherence, analogous to  Jarzynski's and Crooks' relations for work in thermodynamics. Coherence is thus revealed  as another instance of a physical resource, in addition to athermality and entanglement, for which a connection is established between the majorisation framework (regulating pure state transformations under suitable free operations) and the emergence of fluctuation theorems. Our study is hoped to provide further insight into the general structure of battery assisted quantum resource theories, and more specifically into the interplay between quantum coherence and quantum thermodynamics.
\end{abstract}


\section{Introduction}
Quantum coherence is an essential non-classical feature rooted in the foundations of quantum theory. By fixing a particular reference basis $\{|i\rangle\}_{i=1,...,d}$ of the $d$-dimensional Hilbert space $\mathcal{H}$ in which the quantum states of our system of interest live, coherence is simply visualised as the degree to which these states deviate from being diagonal in the chosen basis. Although elementary in its conception, quantum coherence incarnates the essence of superposition and is thus seen as the first step away from a fully classical description of a system, acting as a building block for more advanced phenomena such as entanglement in composite systems.

It is of no surprise therefore that quantum coherence plays a central role in a wide range of quantum technologies, such as metrology, sensing, communication, and imaging. The development of these quantum technologies has motivated the formalisation of quantum coherence as a physical resource within the mathematical framework of resource theories \cite{streltsov2017colloquium}. This has led, {\it inter alia}, to theoretical and experimental investigations of optimal protocols to distill or dilute quantum coherence, and more generally to manipulate and transform quantum states by means of suitably defined free operations unable to create coherence \cite{aaberg2006quantifying,baumgratz2014quantifying,winter2016operational,Streltsov2015b,Napoli2016,Chitambar2016,Chitambar2016b,Chitambar2017,Zhu2017,Wang:17,Wu:17}.


The development of a resource theory of quantum coherence mirrors the early motivation behind the theoretical investigations of classical thermodynamics, where optimal procedures were derived for distilling  work from a thermal machine \cite{carnot1872reflexions}. These have been superseded by the fields of stochastic and quantum thermodynamics \cite{Kosloff,Goold2016}, most notably by the seminal fluctuation theorems due to Jarzynski \cite{jarzynski1997nonequilibrium} and Crooks \cite{crooks1999entropy}, which consider the amount of extractable work as a quantity that can fluctuate during a thermodynamic process, and hence characterise fundamental limitations on the associated work distribution.

Recent work  has formalised a connection between the algebraic theory of majorisation and the emergence of fluctuation theorems \cite{alhambra2016fluctuating, alhambra2017entanglement}. This has been highlighted not only in thermodynamics \cite{alhambra2016fluctuating}, where so-called thermo-majorisation provides necessary and sufficient conditions for state transformations under thermal operations within the resource theory of athermality \cite{horodecki2013fundamental,Gour2013,Goold2016}, but also in the context of entanglement theory \cite{alhambra2017entanglement}, where pure state transformations under local operations and classical communication (LOCC) are once again determined  by majorisation relations \cite{nielsen1999conditions}. These observations raise the prospect that other resources, {\it in primis} coherence, may also be allowed to fluctuate and give rise to a distribution regulated by fluctuation theorems while implementing the conversion of quantum states under the corresponding set of free operations.

In this work we establish fluctuation relations for the manipulation of quantum coherence under incoherent or strictly incoherent operations  \cite{baumgratz2014quantifying,winter2016operational,yadin2016quantum}, that is, another instance where  majorisation theory provides necessary and sufficient conditions for pure state transformations $|\psi\rangle_{A} \rightarrow |\phi\rangle_{A}$ in a quantum system $A$ \cite{winter2016operational,Chitambar2016,Zhu2017}. In order to do this, an ancillary device that stores and supplements coherence is necessary, introduced here as a {\em coherence battery} $B$. This battery, initialised in a state $|\lambda\rangle_{B }$, is used as an approximate catalyst to mediate the pure state transformation $|\Psi\rangle_{AB}\rightarrow|\Phi\rangle_{AB}$ as
  \begin{align}\label{transformation}
  	|\Psi\rangle_{AB}=|\psi\rangle_{A}\otimes|\lambda\rangle_{B}\rightarrow|\Phi\rangle_{AB}\approx|\phi\rangle_{A}\otimes|\lambda\rangle_{B },
  \end{align}\\
where the approximation becomes exact and the transformation reversible in the limit of an ideal battery, as discussed later in Section~\ref{battery}. This establishes a resource-theoretic framework for coherence manipulation under battery assisted incoherent operations (BIO) or battery assisted strictly incoherent operations (BSIO), collectively referred to as B(S)IO.
Necessary and sufficient conditions are derived for the battery to gain or lose a quantity of coherence $w$ probabilistically. This gives rise to a coherence distribution $P(w)$ following the transformation. From this distribution, four theorems characterising fundamental limitations on the manipulation of fluctuating coherence are then derived.

\textbf{Result 1}--- A second law of coherence is derived, which governs the amount of coherence extractable from the battery during the transformation. If coherence is allowed to fluctuate, we find that the average extractable coherence is bounded by the difference in relative entropy between initial and final states of the system.
This complements the fact that the relative entropy of coherence \cite{baumgratz2014quantifying} yields the exact distillable coherence under incoherent operations within the standard resource theory of coherence \cite{winter2016operational}.

\textbf{Result 2}--- A third law of coherence is derived, which demonstrates that the limiting factor for extracting  fluctuating coherence is the diagonal rank of the density matrix. This again mirrors the standard result, namely the rank of the diagonal part of pure state density matrices cannot increase under incoherent operations \cite{winter2016operational}.

\textbf{Result 3}--- An analogue to Jarzynski's relation \cite{jarzynski1997nonequilibrium} is derived, which applies when the final state of the system is maximally coherent. This shows the nature of fluctuating coherence and, in conjunction with the second and third laws of coherence, demonstrates strong bounds on extractable coherence during the transformation.

\textbf{Result 4}--- By comparing forward and reverse transformations, an analogue of Crooks' relation \cite{crooks1999entropy} is found, which applies when the final states of both transformations are maximally coherent. It implies that extracting $w$ units of coherence from the battery in the forward protocol is exponentially suppressed with respect to extracting $-w$ units in the reverse protocol, showing an inherent irreversibility in coherence manipulation.

This paper is organised as follows. Section~\ref{StateTransformation} presents basics of coherence theory and the conditions for state transformations under (strictly) incoherent operations defined by the majorisation criteria. Section~\ref{battery} characterises the coherence battery employed to mediate pure state transformations. Section~\ref{PureTrans} describes necessary and sufficient conditions for battery assisted state transformations, detailing the protocol that gives rise to the fluctuating coherence distribution. Section~\ref{FlucTheorems} presents the aforementioned four results governing the fluctuation relations for coherence. Section~\ref{Discussion} contains a summary and discussion of  our results. Within the Appendix are proofs of the conditions for battery assisted transformations  and the derivation of the reverse protocol necessary for the coherence analogue of Crooks' theorem.

Throughout this work, the density matrix of a pure state $|\psi\rangle_A$ will be denoted by $\psi_A$, the subscript indicating the subsystem to which the state belongs (usually $A$ for the principal system, and $B$ for the battery). The Hilbert spaces of system $A$ and battery $B$ will be denoted by ${\cal H}_A$ and ${\cal H}_B$, and the corresponding set of density matrices by ${\cal D}({\cal H}_A)$ and  ${\cal D}({\cal H}_B)$, respectively. Occasionally subsystem labels will be omitted when clear from the context.

\section{State transformations in the resource theory of coherence} \label{StateTransformation}
Let us begin by briefly reminding the basics of quantum coherence from a resource theoretic perspective, referring the reader to \cite{streltsov2017colloquium} for more details.

Given a reference basis $\{|i\rangle\}_{i=1,...,d}$ of a $d$-dimensional Hilbert space $\mathcal{H}$, such as e.g.~the computational basis, density matrices of the form
\begin{equation}\label{incoherentstates}
\sigma = \sum_i c_i |i\rangle\langle i|\,,
\end{equation}
 form the set $\mathcal{I}$ of incoherent states. For a composite system with Hilbert space $\mathcal{H}_{AB} = \mathcal{H}_A \otimes \mathcal{H}_B$, the reference basis is taken as the tensor product of the reference bases of each individual subsystem, and the set $\mathcal{I}$ of incoherent states is defined accordingly.

Incoherent operations (IO) are completely positive trace preserving maps $\Lambda$ which admit an operator sum representation such that all Kraus operators $\{K_l\}$ map incoherent states into incoherent states, that is,
$\Lambda (\rho) = \sum_l K_l \rho K_l^{\dagger}$, with $\sum_{l}K_{l}^{\dagger}K_{l}=\mathds{1}$ and
\begin{equation}\label{IO}
\frac{K_{l}\sigma K_{l}^{\dagger}}{\text{Tr}(K_{l}\sigma K_{l}^{\dagger})} \in  {\cal I}\,, \quad \forall\, \sigma \in {\cal I}\,.
\end{equation}
This definition entails that IO cannot create coherence from an incoherent state, not even probabilistically.

Strictly incoherent operations (SIO) are a subclass of IO whose Kraus operators additionally satisfy \cite{winter2016operational}
\begin{equation}\label{SIO}
\frac{K_{l}^{\dagger}\sigma K_{l}}{\text{Tr}(K_{l}^{\dagger}\sigma K_{l})} \in  {\cal I}\,, \quad \forall\, \sigma \in {\cal I}\,.
\end{equation}
This equivalently means that the results of measuring (in the reference basis) an output state after SIO do not depend on the coherence of the input state $\rho$ \cite{yadin2016quantum},
\begin{equation}
\langle i|K_{l}\rho K_{l}^{\dagger}|i\rangle=\langle i|K_{l}\Delta(\rho)K_{l}^{\dagger}|i\rangle\,,
\end{equation}
where we have introduced the dephasing operation $\Delta$, whose action is defined as
\begin{equation}\label{delta}
\Delta(\rho) \coloneqq \sum_{i=1}^{d}|i\rangle\langle i|\rho|i\rangle\langle i|\,.
\end{equation}
\label{eq:dephasing}

There are several monotones apt to quantify the degree of coherence of a quantum state $\rho$ \cite{streltsov2017colloquium}. We will adopt the relative entropy of coherence $C_{\text{rel}}$ \cite{aaberg2006quantifying,baumgratz2014quantifying,Herbut2005,Vaccaro2008,Gour2009}, which takes the simple closed form
\begin{equation}\label{Crel}
C_{\text{rel}}(\rho)\coloneqq S(\Delta(\rho))-S(\rho)\,,
\end{equation}
where $S(\rho) \coloneqq -\text{Tr}(\rho \ln \rho)$ is the conventional von Neumann entropy, that is used prominently in quantum information theory as well as in extensive thermodynamics. The relative entropy of coherence admits a valuable operational interpretation as it amounts to the distillable coherence under IO in an asymptotic setting
\cite{winter2016operational} \footnote{The distillable coherence $C_{\text{d}}(\rho)$ of a state $\rho$ is defined as the maximum ratio $R$ such that the conversion $\rho^{\otimes m} \rightarrow  \ket{\phi^+_2}\bra{\phi^+_2}^{\otimes mR}$ can be implemented by IO in the limit of many copies $m \rightarrow \infty$. Strictly speaking, the equality between distillable coherence and relative entropy of coherence holds when $\log_2$ is used instead of $\ln$ in the definition of  entropy, as it is customary in information theory. In this paper, we adopt instead natural logarithms to better emphasise the connection with thermodynamics, which means that in our notation we have $C_{\text{d}}(\rho)=C_{\text{rel}}(\rho)/\ln(2)$.}.

A maximally coherent state in a Hilbert space of dimension $d$ can be written as a uniform superposition of the reference basis states,
\begin{equation}\label{maxcoh}
\ket{\phi^+_d} \coloneqq \frac{1}{\sqrt{d}} \sum_{i=1}^d \ket{i}\,,
\end{equation}
and its coherence is given by $C_{\text{rel}}(\phi_d^+)=\ln (d)$.

\begin{figure} [tbh]
	\begin{center}
		\includegraphics[width=\columnwidth]{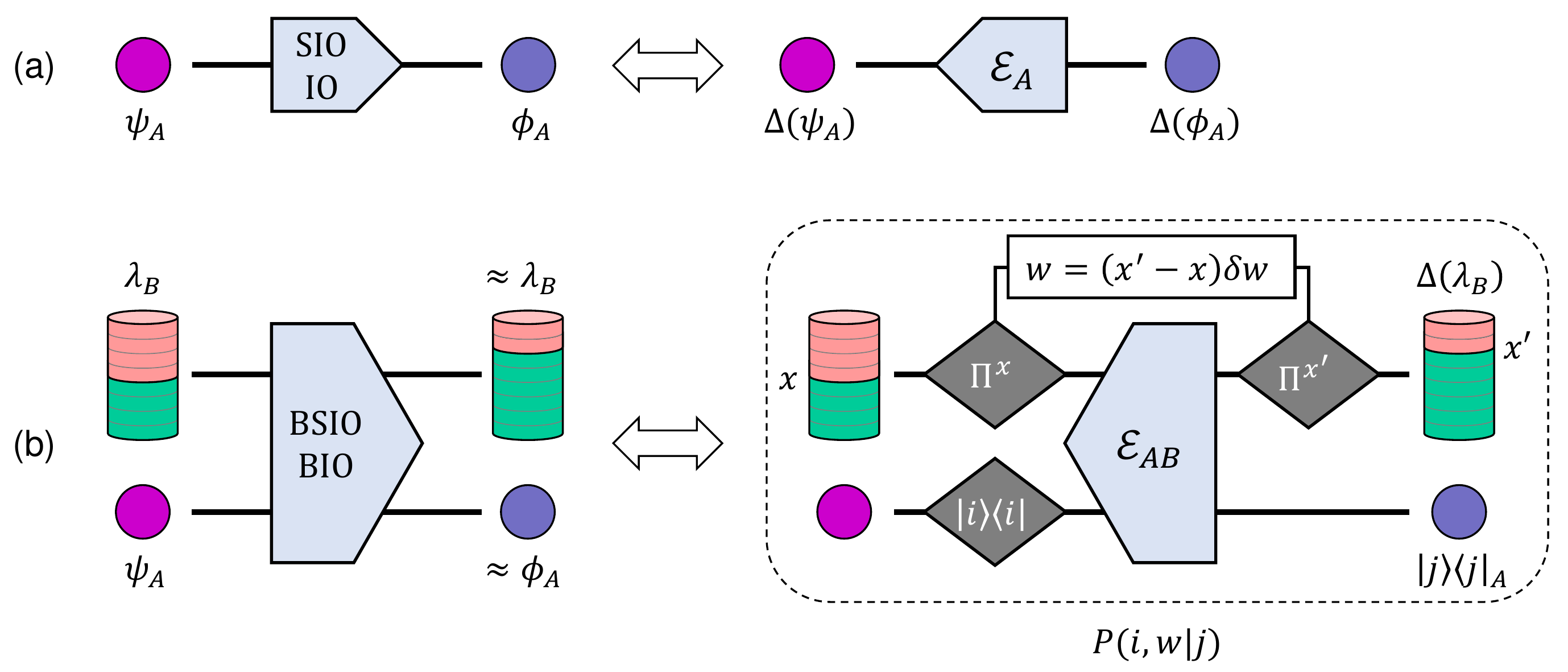}
		\caption{{\bf (a)} In the standard resource theory of quantum coherence \cite{streltsov2017colloquium}, the pure to pure state transformation $\ket{\psi}_A \rightarrow \ket{\phi}_A$ on the system $A$ (left) can be implemented by SIO or IO if and only if there exists a unital map $\EE_A$ of the form (\ref{map}) that maps the diagonal component of the final state $\Delta(\phi_A)$ into that of the initial state $\Delta(\psi_A)$ (right). {\bf (b)} In the battery assisted framework considered here, the pure to pure state transformation  $\ket{\psi}_A  \otimes \ket{\lambda}_B \rightarrow \ket{\Phi}_{AB} \approx \ket{\phi}_A \otimes \ket{\lambda}_B$ (left)  can be implemented by BSIO or BIO --- that is, by SIO or IO on the system $A$ and the battery $B$, accompanied by a change in coherence of the battery by an amount $w$ with probability $P(w)$  --- if and only if there exists a conditional probability distribution $P(i,w|j)$ that satisfies the three conditions given in Eqs.~(\ref{Condition 1})--(\ref{Condition 3}). Such a  distribution can be constructed from the statistics of the  protocol illustrated in the dashed box (right),  described in Section~\ref{PureTrans}. This framework allows us to investigate fluctuation relations for quantum coherence, analogous to those for work in thermodynamics, as presented in Section~\ref{FlucTheorems}. \label{fig:circuit}	}
	\end{center}
\end{figure}

In this paper we consider a pure to pure state transformation
\begin{equation}\label{pst}
	\ket{\Psi}\rightarrow\ket{\Phi}\,,
\end{equation}
It is known that such a transformation is possible by means of general deterministic SIO or IO, that is, $\exists\, \Lambda \in \text{(S)IO}$ such that $\Phi = \Lambda(\Psi)$,  if and only if \cite{Du2015b,Du2017,winter2016operational,Chitambar2016,Chitambar2016b,Chitambar2017,Zhu2017}
\begin{equation}\label{majo}
\Delta(\Psi) \prec \Delta(\Phi)\,,
\end{equation}  that is, $\Delta(\Psi)$ is majorised by $\Delta(\Phi)$. Explicitly, the necessary and sufficient condition for this majorisation relation to hold is \cite{bhatia_1996,nielsen_2001}
\begin{equation}\begin{aligned}\label{trans}
\Delta(\Psi) = \sum_m r_m \Xi_m \Delta(\Phi) \Xi^\dagger_m\,,
\end{aligned}\end{equation}
where $r_m \geq 0,\; \sum_m r_m = 1$, and $\Xi_m$ are permutation matrices. Defining this in terms of a completely positive trace preserving unital map $\EE$ acting on any operator $X$,
\begin{equation}\begin{aligned}\label{map}
\EE(X) \coloneqq  \sum_m r_m \Xi_m X \Xi^\dagger_m\,,
\end{aligned}\end{equation}
we then have that, as depicted in Figure~\ref{fig:circuit}{(a)}, the pure state transformation in Eq.~(\ref{pst}) can be implemented by (S)IO if and only if there exists a unital map $\EE$ of the form (\ref{map}) such that
\begin{equation}\label{majomap}\Delta(\Psi) = \EE(\Delta(\Phi))\,.
\end{equation}
This is equivalent to the existence of a bistochastic matrix mapping the (nonzero) diagonal coefficients of $\Phi$ to those of $\Psi$ \cite{bhatia_1996}.

\section{Coherence battery}\label{battery}
We consider a system $A$ on which we aim to perform the pure to pure state transformation $\ket{\psi}_A \rightarrow \ket{\phi}_A$, supplemented by a battery $B$, so that the composite state transformation can be written overall as in Eq.~(\ref{transformation}). The battery is initialised in a state $|\lambda\rangle_{B }$ that can be defined in general as a superposition of coherence eigenstates  $|c_x\rangle$,
\begin{equation}\label{batstatefull}
|\lambda\rangle_{B }\coloneqq \sum_{x=0}^{n}\sqrt{\alpha_x}|c_x\rangle_{B }\,,
\end{equation}
with $\alpha_x \geq 0$, $\sum_x \alpha_x = 1$.
Here by coherence eigenstates we mean states $|c_x\rangle$ with a well defined amount of coherence, as quantified by the relative entropy $C_{\text{rel}}$. In particular, in analogy  to the case of the entanglement battery studied in \cite{alhambra2017entanglement}, we can write each $|c_x\rangle$ as the tensor product of two types of states, namely $x$ copies of a state $|\Upsilon^{+}_{u}\rangle$ with higher coherence (i.e., a charged state) and $n-x$ copies of a state $|\Upsilon^{-}_{u}\rangle$ with lower coherence (i.e., a discharged state). Precisely,
\begin{align}\label{batstate}
|c_{x}\rangle \coloneqq \underbrace{|\Upsilon^{+}_{u}\rangle\otimes....\otimes |\Upsilon^{+}_{u}\rangle}_{x}\,\otimes\,\underbrace{|\Upsilon^{-}_{u}\rangle\otimes.....\otimes|\Upsilon^{-}_{u}\rangle}_{n-x},
\end{align}
for chosen integers $n$ and $u$ and for all integers  $x \in \{0,....,n\}$, with
\begin{align}
|\Upsilon^{+}_{u}\rangle&\coloneqq \frac{1}{\sqrt{u}}\sum_{i=1}^{u}|i\rangle\,, \label{+} \\
|\Upsilon^{-}_{u}\rangle&\coloneqq \frac{1}{\sqrt{u-1}}\sum_{i=u+1}^{2u-1}|i\rangle\,. \label{-}
\end{align}
From Eq.~(\ref{Crel}), we see that the relative entropy of coherence of the charged and discharged states is given respectively by
\begin{align}
&C_{\text{rel}}({\Upsilon^{+}_u})=\ln(u),\\
&C_{\text{rel}}({\Upsilon^{-}_u})=\ln(u-1).
\end{align}
Note that the states $|\Upsilon^{+}_{u}\rangle$ and $|\Upsilon^{-}_{u}\rangle$ are equivalent to maximally coherent states of dimension $u$ and $u-1$, respectively.
The coherence of the state (\ref{batstate}) is then given by
\begin{equation}\label{Ccx}
C_{\text{rel}}(c_x)=\ln\left(u^x (u-1)^{n-x}\right)\,.
\end{equation}

A measurement of the `position' of the battery, or more properly, of its level of coherence as specified by the index $x$, can be obtained by defining a set of orthogonal projectors $\Pi^x$ as
\begin{align}\label{proj}
\Pi^x\coloneqq u^x(u-1)^{n-x}\chi_x\,,
\end{align}
where $\chi_x$ is the incoherent state corresponding to the diagonal part of the state (\ref{batstate}),
\begin{align}\label{redAlic}
\chi_x\coloneqq \Delta(|c_x\rangle\langle c_x|)&=\Big{(}\frac{1}{u}\sum_{i=1}^{u}|i\rangle\langle i|\Big{)}^{\otimes x}\otimes \Big{(}\frac{1}{u-1}\sum_{i=u+1}^{2u-1}|i\rangle\langle i|\Big{)}^{\otimes(n-x)}\,.
\end{align}
By construction, $\sum_x \Pi^x =: \mathds{B}_B$, which is the projector on the diagonal support of the battery; in other words,
the projectors $\{\Pi^x\}$ give a resolution of the identity on the subspace ${\cal B} \subset {\cal D}({\cal H}_B)$ of the state space of the battery,  spanned by $\text{supp}\big(\Delta(\lambda_B)\big)=\bigoplus_{x=0}^n \text{supp}(\chi_x)$.

Similarly to a conventional energy storage device, the battery $B$ will act as a coherence supplier that can receive/transfer coherence from/to the system $A$, by changing the ratio of the states  \eqref{+} and \eqref{-}. In fact, the discharging process $|\Upsilon^{+}_{u}\rangle \rightarrow |\Upsilon^{-}_{u}\rangle$ corresponds to decreasing $x$ by one and hence diminishes the coherence in the battery by a {\it quid} $\delta w$,
\begin{align}\label{deltaw}
\delta w \coloneqq \ln\left(\frac{u}{u-1}\right)\,.
\end{align}
This can be seen as extracting one unit of coherence from the battery.

In general, the role of the battery is to mediate the state transformation (\ref{transformation}) by exchanging an amount $w$ of coherence with the system. We may choose the parameter $u$ large enough, corresponding to a level spacing $\delta w \approx 1/u$ in the battery fine enough, so that any change $w$ in the coherence of the battery can be taken approximately to be a multiple of $\delta w$. Ideally, we would like the battery to be reusable in order to assist subsequent state transformations. Furthermore, we would like the battery to serve the purpose of overcoming the limitations in conventional (unassisted) state transformations under (S)IO on the system, going beyond the conditions of Section~\ref{StateTransformation}.
Therefore, there are three constraints that an ideal battery should adhere to:
\begin{itemize}
	\item[1.] In order for the final state $|\phi\rangle_{A }$ to be pure, the system should be virtually uncorrelated with the battery, $|\Phi\rangle_{AB}\approx|\phi\rangle_A\otimes|\lambda\rangle_{B }$.
	\item[2.] The only allowed action on the battery should be the raising and lowering of $w$ units of coherence, by the unitary operator $\Gamma^w$ defined as \begin{equation}\label{raising}
\Gamma^{w}|c_{x}\rangle_{B }=|c_{x+w}\rangle_{B }\,,
\end{equation}
with $x+w$ assumed modulo $n+1$, and $x$ and $n$ assumed large enough to avoid hitting the bottom or top levels of the battery.
	\item[3.] The state of the battery $|\lambda \rangle_{B }$ should allow for approximately implementing all reversible pure to pure state transformations $|\psi\rangle_{A }\rightarrow|\phi\rangle_{A }$ by (S)IO.
\end{itemize}

The first constraint is fulfilled provided the chosen battery state  $|\lambda\rangle_B$ is a superposition over sufficiently many eigenstates $|c_{x}\rangle$, that is, provided the size of the battery, determined by the parameter $n$, is chosen large enough.
The last two constraints are  stronger and force the state of the battery $|\lambda\rangle_B$ to be close to a uniform superposition of coherence eigenstates $|c_{x}\rangle$. To see this, note that the second constraint imposes that the final state of system and battery has to be of the form
\begin{align}\label{final_sum}
|\Phi \rangle_{AB}=\sum_{w:|w| \leq w_{\max}}|\phi_w\rangle_{A }\otimes \Gamma^{w}|\lambda\rangle_{B }\,,
\end{align}
for some $w_{\max} >0$,
while the third constraint imposes that, for all reversible transformations (\ref{transformation})
implemented by (S)IO, the final state is  $\epsilon$-close to the target one,
\begin{align}\label{close}
||\Phi_{AB}-\phi_{A}\otimes\lambda_{B}||_{1}\leq\epsilon\,,
\end{align}
and with identical diagonal marginal on the system,
\begin{align}\label{marginal}
\Delta(\text{Tr}_B \Phi_{AB}) =\Delta(\phi_A)\,.
\end{align}

As the conditions above have to hold for {\it all} reversible state transformations, we can analyse the specific one where the initial and final states of the system are, respectively, \begin{align}
|\psi\rangle_{A}&=\frac{1}{\sqrt{2}}(|0\rangle_{A }+|\Upsilon^+_u\rangle_{A})\,,\label{eg_state1}\\
|\phi\rangle_{A}&=\frac{1}{\sqrt{2}}(|0\rangle_{A}+|\Upsilon^-_u\rangle_{A})\,.\label{eg_state2}
\end{align}
Considering now system and battery initialised in the state $\ket{\Psi}_{AB} = \ket{\psi}_A \otimes \ket{\lambda}_B$, and noting that the transformation $\ket{\Psi}_{AB} \rightarrow \ket{\Phi}_{AB}$ can be implemented reversibly under SIO \cite{winter2016operational} or IO \cite{Chitambar2016,Zhu2017} if and only if the nonzero diagonal coefficients of the initial and final states are identical, one can show that the only final state of system and battery that fulfils this requirement while complying with Eqs.~(\ref{final_sum}) and (\ref{marginal}) is
\begin{align}\label{final}
|\Phi\rangle_{AB}=\frac{1}{\sqrt{2}}(|0\rangle_{A }\otimes|\lambda\rangle_{B }+|\Upsilon^{-}_u\rangle_{A }\otimes\Gamma^{\delta w}|\lambda\rangle_{B }).
\end{align}
The proof follows closely the one reported in \cite{alhambra2017entanglement} for LOCC transformations in entanglement theory, with diagonal coefficients here playing the same role as Schmidt coefficients there. Finally invoking Eq.~(\ref{close}) and applying further algebra \cite{alhambra2017entanglement}, one finds
\begin{align}\label{limit}
\sum_x |\alpha_x -\alpha_{x+y}|\leq |y| \sqrt{8\epsilon}\,, \quad \forall |y| \leq w_{\max}/\delta w\,.
\end{align}
This means that, in order for the battery to serve as an approximate catalyst to implement all reversible (S)IO pure to pure state transformations, including the specific instance just discussed, the set of coefficients $\alpha_x$ in its initial superposition state $\ket{\lambda}_B$ of the form (\ref{batstatefull}) must be close to a uniform distribution, as formalised by Eq.~(\ref{limit}).

\section{Battery assisted state transformations and coherence distribution protocol} \label{PureTrans}

We are now ready to investigate necessary and sufficient conditions for the  transformation between initial and final states $\ket{\psi}_A$ and $\ket{\phi}_A$ of the system $A$, with diagonal components $\Delta(\psi_A)=\sum_i p_i \ket{i}\bra{i}_A$ and $\Delta(\phi_A)=\sum_j q_j \ket{j}\bra{j}_A$, mediated by a change in coherence of an amount $w$ with probability distribution $P(w)$ in the battery $B$, initially prepared in the state $\ket{\lambda}_B$ of Eq.~(\ref{batstatefull}).
Here the distribution $P(w)$ is associated to a two-stage measurement of the battery with the projectors (\ref{proj}) before and after the transformation, that is, $P(w)$ is the probability of finding the battery in the final state $\ket{c_{x+w}}_B$, given that it was found initially in the state $\ket{c_x}_B$.

 The main result of this Section, which mirrors the analogous one recently reported for entanglement theory \cite{alhambra2017entanglement},
is illustrated in Figure~\ref{fig:circuit}{(b)} and can be enunciated as follows.
\begin{thm*}[Necessary and sufficient conditions for B(S)IO transformations]\label{thm0}
The transformation of Eq.~(\ref{transformation}) can be implemented by means of SIO or IO on the system and the battery while extracting a coherence distribution $P(w)$ --- that is, by battery assisted (S)IO or, in short, B(S)IO --- if and only if there exists a conditional probability distribution $P(i,w|j)$, with marginals $P(i)=p_i$, $P(j)=q_j$,  and $P(w)$, which fulfils the following three conditions:
\begin{align}
&\text{Condition 1:}\,\,\,	\sum_{i,w}P(i,w|j)=1,\,\,\, \forall j\,,\label{Condition 1}\\
&\text{Condition 2:}\,\,\, \sum_{j,w}P(i,w|j)e^w=1, \,\,\, \forall i\,,\label{Condition 2}\\
&\text{Condition 3:}\,\,\, \sum_{j,w} P(i,w|j)q_j=p_i,\,\,\, \forall i\,.\label{Condition 3}
\end{align}
\end{thm*}

Physically, Condition 1 expresses the normalisation of the conditional probability distribution $P(i,w|j)$, Condition 2 regulates the fluctuations of $w$ units of coherence in the battery with probability $P(w)$, while Condition 3 formalises the requirement that the marginals $P(i)$ and $P(j)$ of the joint probability distribution $P(i,j,w)=P(i,w|j)P(j)$ reproduce the diagonal components $p_i$ and $q_j$ of the initial and final states of the system, respectively.

\begin{proof}The proof of the Theorem consists of two directions. For the ``if'' part,
we need to show that, given a conditional probability distribution $P(i,w|j)$ obeying Conditions 1--3, sequences of B(S)IO protocols $\Lambda^{(N)}_{AB}$ and states $\ket{\Psi^{(N)}}_{AB}$ and $\ket{\Phi^{(N)}}_{AB}$ exist, such that $\Phi^{(N)}_{AB}=\Lambda^{(N)}_{AB}(\Psi^{(N)}_{AB})$, with $\lim_{N \rightarrow \infty} \ket{\Psi^{(N)}}_{AB}=\ket{\psi}_A \otimes \ket{\lambda}_B$ and $\lim_{N \rightarrow \infty} \ket{\Phi^{(N)}}_{AB}=\ket{\phi}_A \otimes \ket{\lambda}_B$. As recalled in Section~\ref{StateTransformation}, this is equivalent to showing the existence of a sequence of bistochastic matrices $G^{(N)}$ mapping the (nonzero) diagonal coefficients of $\Phi^{(N)}_{AB}$ to those of $\Psi^{(N)}_{AB}$. Such a derivation is rather technical and hence deferred to~\ref{A1}.

For the ``only if'' part, let us assume that a B(S)IO transformation (\ref{transformation}) is possible, that is, there exists a unital map $\EE_{AB}$ of the form \eqref{map} such that the (nonzero) diagonal components of the initial and final states satisfy
\begin{equation}\label{EEab}
\Delta(\Psi_{AB})=\EE_{AB}(\Delta(\Phi_{AB}))\,.
\end{equation}
We then  need to prove that a conditional probability distribution $P(i,w|j)$ fulfilling the above three conditions exists. It turns out one can explicitly construct such a probability distribution from the following five step protocol, also schematically represented in Figure~\ref{fig:circuit}{(b)}:
\begin{itemize}
	\item[1.] Prepare the incoherent state $|j \rangle \langle j|_{A} \otimes \Delta(\lambda_{B})$;
	\item[2.] Measure the battery with the projector $\Pi_B^{x'}$ from \eqref{proj};
	\item[3.] Transform the resulting state of system and battery with the unital map $\mathcal{E}_{AB}$ of Eq.~(\ref{EEab});
	\item[4.] Measure the system with the projector $|i \rangle \langle i|_{A}$ and the battery with  $\Pi_B^x \equiv \Pi_B^{x'-\frac{w}{\delta w}}$;
	\item[5.] Record the variable $w=(x'-x)\delta w$, discarding $x$ and $x'$.
\end{itemize}

The protocol above, in which  $\frac{w}{\delta w}$ describes the amount $w$ of extracted coherence as a multiple of the unit $\delta w$ defined in \eqref{deltaw}, gives rise to the probability distribution
\begin{align}\label{con_prob}
P(i,w|j)&=\sum_{x'}\text{Tr}[(|i\rangle \langle i |_A \otimes \Pi^{x'-\frac{w}{\delta w}}_B)\mathcal{E}_{AB}(|j\rangle \langle j|_A \otimes \Pi^{x'}_B\Delta(\lambda_{B})\Pi^{x'}_B)]\,,
\end{align}
which, using Eq.~\eqref{proj}, can also be rewritten as
\begin{align}\label{con_prob2}
P(i,w|j)=\sum_{x'}\alpha_{x'}\text{Tr}[(|i\rangle \langle i |_A \otimes \Pi^{x'-\frac{w}{\delta w}}_{B})\mathcal{E}_{AB}(|j\rangle \langle j|_A \otimes {\chi_{x'}}_{B})].
\end{align}
The proof that $P(i,w|j)$ satisfies Conditions 1--3 is reported in~\ref{A2}.
\end{proof}

\section{Fluctuation theorems from coherence distribution}\label{FlucTheorems}
We have shown that the amount $w$ of fluctuating coherence exchanged between battery and system when mediating a pure to pure state transformation $\ket{\psi}_A \rightarrow \ket{\phi}_A$ gives rise to a conditional probability distribution \eqref{con_prob2}. In analogy to the derivation of the  fluctuation theorems from a conditional work probability distribution in thermodynamics \cite{crooks1999entropy},  several coherence fluctuation theorems can now be obtained. In this Section, we present the mathematical derivation of the four main results anticipated in the Introduction, accompanying each of them with relevant physical remarks and comparisons with the corresponding thermodynamic laws.

\subsection{Second law of coherence}
\noindent\textit{The following is for an initial state $|\psi\rangle_A$ with diagonal coefficients $p_i$ and a target state $|\phi\rangle_A$ with diagonal coefficients $q_j$.}\\

Starting with Condition 2, multiplying Eq.~(\ref{Condition 2}) by $p_i$, rewriting the conditional probability distribution as $P(i,w|j)=P(i,j,w)/q_j$, and summing over $i$ gives
\begin{align}
\sum_{i,j,w}P(i,j,w)\frac{p_i}{q_j}e^w=1\,,
\end{align}
where we have used Condition 3. Now using $\frac{p_i}{q_j}=e^{\ln \big(\frac{p_i}{q_j}\big)}$  to move the probabilities into the exponent,
\begin{align}
\sum_{i,j,w}P(i,j,w)e^{w-\ln q_j+\ln p_i}=1\,,
\end{align}
and writing in bracket form, we get
\begin{align}\label{unexpanded}
\langle e^{w-\ln q_j+\ln p_i}\rangle=1\,.
\end{align}
This describes the distribution of fluctuating coherence $w$ that can be extracted during our pure state transformation. By expanding to first order, and using Eq.~(\ref{Crel}), we find
\begin{align}\label{first}
	\langle w \rangle \leq C_{\text{rel}}(\psi_{A})-C_{\text{rel}}(\phi_{A})\,.
\end{align}
This shows that the \textit{average} coherence extractable from the battery to mediate B(S)IO state transformations is bounded by the difference in relative entropy of coherence between the initial and final states of the system. This is in contrast to the standard operational setting in the resource theory of quantum coherence, in which the relative entropy of coherence (scaled by a factor $\ln(2)$ in our notation)  quantifies the exact distillable coherence under IO \cite{winter2016operational}.

 We can also see that Eq.~(\ref{first}) is formally analogous to the traditional second law of thermodynamics, $\langle W \rangle \le F(\rho)-F(\sigma)$, which states that during the state transformation $\rho\rightarrow\sigma$ the average work $W$ required is less than or equal to the difference in free energies $F(\rho)=\langle H\rangle-TS(\rho)$. As Eq.~(\ref{first}) is the first order expansion, this is just the average result. Higher order Taylor expansions of Eq.~\eqref{unexpanded} lead to all the moments of the coherence distribution $P(w)$ that can be obtained during the transformation.

\subsection{Third law of coherence}
\noindent\textit{The following is for the transformation $|\psi\rangle_A  \rightarrow|\phi\rangle_A$, where $p_{\text{min}}$ and $q_{\text{min}}$ are the smallest nonzero diagonal coefficients of the initial and final state, respectively.}\\

Starting with Condition 3 on the probability distribution \eqref{Condition 3},
\begin{align}
\sum_{j,w} P(i_0,w|j)q_j=p_{\text{min}}\,,
\end{align}
where $i_0$ is the index corresponding to the smallest diagonal coefficient of the state,  $p_{i_0} \equiv p_{\text{min}}$, we can write the inequality
$P(i_0,w|j)q_j\le p_{\text{min}}\,\,\,\forall j$, that can be substituted into Eq.~\eqref{Condition 2}, to get
\begin{align}
1 \leq \sum_{j,w}\frac{p_{\text{min}}}{q_j}e^w \leq  \sum_{j,w}\frac{p_{\text{min}}}{q_{\text{min}}}e^w\,,
\end{align}
where we further used the fact that $1/q_j\le 1/q_{\text{min}}\,\,\,\forall j$ by construction.
Summing the rightmost term over $j$, the diagonal rank  $d'$ of the final state is obtained, so that we can write
\begin{align}\label{third}
\sum_we^w\ge\frac{q_{\text{min}}}{d'p_{\text{min}}}\,.
\end{align}
This can be interpreted as  the third law for fluctuating coherence. It is well known that the majorisation criterion (\ref{majo}) for state transformations in the resource theory of coherence implies the following statement, namely that the rank of the diagonal part of pure states cannot increase under (S)IO \cite{winter2016operational}. Here, we find that the amount of fluctuating coherence $w$ required to increase the diagonal rank (i.e., to send $p_{\text{min}} \rightarrow 0$) under B(S)IO must diverge. Therefore such an operation is forbidden as it would require a battery of infinite size. The analogous result in thermodynamics is that decreasing the rank of a density matrix requires infinite resources, which can be regarded as a general  statement of the third law \cite{masanes2017general}.


\subsection{Jarzynski's relation for coherence}\label{secJJ}
\noindent\textit{The following is for the transformation $|\psi\rangle_A  \rightarrow|\phi\rangle_A$, where the final state is a maximally coherent state of dimension $d'$. }\\

Starting again from Condition 2, and using the fact that $q_j=1/d'$ $\forall j$ for a maximally coherent final state, we can multiply both sides of Eq.~(\ref{Condition 2}) by $1/d'$  and obtain
\begin{align}
\sum_{j,w}P(i,j,w)e^w=\frac{1}{d'}\,.
\end{align}
Now summing over the index $i$ gives
\begin{align}
\sum_{i,j,w}P(i,j,w)e^w=\sum_{i}\frac{1}{d'}=\frac{d}{d'}\,,
\end{align}
which can again be written in bracket form,
\begin{align}\label{Jarzynski}
\langle e^{w} \rangle=\frac{d}{d'}\,,
\end{align}
where $d$ and $d'$ denote the diagonal rank of the initial and final states, respectively.

In statistical mechanics, Jarzynski's relation \cite{jarzynski1997nonequilibrium}  $\langle e^{\beta W} \rangle=\frac{Z'}{Z}$ describes an initial thermal state which is driven out of equilibrium to a final state with a different Hamiltonian $H'$, where $Z$ and $Z'$ are the initial and final partition functions, respectively. Jarzynski's relation implies that, when trying to extract work from a thermal bath, the probability of success decreases exponentially with the amount of work $W$ being extracted.
Equation~\eqref{Jarzynski} equivalently says that attempting to extract more coherence than the average for an initial state of dimension $d$ results in the dimension of the maximally coherent final state, $d'$, decreasing. As known by the analogous third law of coherence in Eq.~\eqref{third}, decreasing the dimension of the final state is limited by the maximum fluctuating coherence $w$ and thus by the size of the battery. Explicitly, from Eq.~(\ref{Jarzynski}) it follows that $P\big(w \geq \ln (\frac{d}{d'})+r\big) \leq e^{-r}$.

The comparison between Eq.~(\ref{Jarzynski}) and  Jarzynski's equation also highlights an analogy between $d/d'$ and $Z'/Z$. Using the relation from statistical mechanics $F=-1/\beta \ln (Z)$, where $F$ is the free energy of the state, the ratio $Z'/Z=e^{\beta (F-F')}$ is describing the exponential of the extractable work from the state under thermal operations. Similarly for $d/d'$, using the relation $C_{\text{rel}}=\ln d$ for a pure (maximally coherent) state, the ratio $d/d'=e^{C_{\text{rel}}'-C_{\text{rel}}}$ is expressing the exponential of the extractable coherence from the state under B(S)IO. Note also that the fluctuation relation in Eq.~(\ref{Jarzynski}) holds for a whole family of (not necessarily maximally coherent) initial states with the same $d$; this is mirrored by the redefinition of free energy from a single average value to a family of free energies with the same fluctuating behaviour \cite{alhambra2016fluctuating}, giving rise to the `many second laws' of quantum thermodynamics \cite{Brandao2015a}.

\subsection{Crooks' relation for coherence}
\noindent\textit{The following is for the transformation $|\psi\rangle_A  \rightarrow|\phi\rangle_A$ and its reverse $|\psi'\rangle_A  \rightarrow|\phi'\rangle_A$,
where the final states of the forward and reverse transformations are maximally coherent states of dimension $d'$ and $d$ respectively.}\\

Crooks' theorem of statistical mechanics \cite{crooks1999entropy} relates the forward and reverse probabilities of extracting a quantity of work $W$ during a non-equilibrium transformation between two thermal states,
\begin{equation}\label{Croods}
\frac{P(W)}{P^{\text{rev}}(-W)}=e^{-\beta  W } \frac{Z'}{Z}\,.
\end{equation}
Crooks' equation shows that the forward protocol is exponentially suppressed in comparison to its reverse. This is a quantitative description of the emergent irreversibility of thermodynamics.

In order to find the coherence analogue of this relation, a reverse transformation protocol is derived in~\ref{Crooks}, where an integer quantity of coherence $-\frac{w}{\delta w}$ is extracted. This results in forward and reverse coherence distributions of the form
\begin{align}
P(w)=&\sum_{i,j}P(i,w|j)\frac{1}{d'}\,,\\
P^{\text{rev}}(-w)=&\sum_{i,j}P^{\text{rev}}(j,-w|i)\frac{1}{d}\,.
\end{align}
It is shown in~\ref{Crooks} that these distributions obey the relation
\begin{align}\label{CohCrooks}
\frac{P(w)}{P^{\text{rev}}(-w)}=e^{-w}\frac{d}{d'}\,.
\end{align}
  In analogy to Crooks' theorem it can be seen that extracting $w$ units of coherence in the forward protocol is exponentially less likely than extracting $-w$ in the reverse protocol. One could attempt to increase the preference of the forward protocol $P(w)$ by decreasing the diagonal rank $d'$ of its final state, however according to the equivalent third law of coherence in Eq.~\eqref{third} this is exponentially difficult in its own right. It has therefore been shown that there is an inherent irreversibility in the manipulation of coherence within the B(S)IO framework.

\section{Discussion}\label{Discussion}

In this work we have established fluctuation relations for the manipulation of quantum coherence, in the context of pure to pure state transformations via (strictly) incoherent operations, assisted by a coherence battery which can exchange coherence with the system probabilistically. We hope that this work motivates a further reconsideration of coherence in quantum mechanics, from a useful albeit static resource that may be invested to convert quantum states and realise useful technological applications,  to a quantity that more generally can fluctuate during transformations and may enable otherwise impossible applications. This has been accomplished here, similarly to the way in which work has been redefined in quantum thermodynamics \cite{alhambra2016fluctuating} and more recently entanglement has been investigated as a fluctuating quantity in quantum information theory \cite{alhambra2017entanglement}.  It is surprising that by forming parallels between coherence, entanglement, and work distribution protocols, these three seemingly disparate quantities obey formally analogous fluctuation theorems.

The obvious underlying link between the state transformations in these distinct contexts is the central role played by the majorisation criteria in the corresponding resource theories. This leads to the primary open question of whether majorisation is necessary or just sufficient for the emergence of fluctuation theorems. The present investigation also suggests that there may be more operational contexts in quantum mechanics and beyond for which specific resources can be considered to fluctuate. In this respect, a fascinating problem is whether one might establish a hierarchy of fluctuating resources, whereby some fundamental quantity which obeys fluctuation theorems can be shown to induce a similar behaviour onto other resources. Quantum coherence is in fact an essential ingredient not only for entanglement, but also for more general non-classical correlations \cite{streltsov2017colloquium,ABC}. It could be worthwhile to address whether fluctuation relations for the latter can be suitably derived by adapting and extending the present work.
We further remark that all these phenomena can be naturally quantified by means of extensive quantities, based on the von Neumann and R\'enyi entropies, but generalisations of our results to non-extensive settings \cite{nonextensive}, e.g.~adopting quantifiers based on Tsallis entropies, could also potentially be considered.

It is intriguing that an ancillary system to transfer and store coherence is necessary in the protocol we introduced, just like a conventional battery (a system that is able to store and transfer work, such as a weight or a piston) is used in thermodynamics. However, unlike work in classical thermodynamics, which can be determined by just measuring the difference in energy between the initial and final state, here to extract the distribution of fluctuating coherence e.g.~according to the measurement protocol in Figure.~\ref{fig:circuit}{(b)}, one irremediably destroys the initial superposition. It is this very uncertainty associated to fluctuations of coherence in the battery that ultimately allows the implementation of state transformations which would otherwise be forbidden to occur.

Further comparisons can be made between the fields.
Because of the definition of the majorisation criteria for the state transformation (\ref{transformation}), which map the diagonal elements of the initial state to those of the final state (rather than vice versa), the probability distribution in Eq.~\eqref{con_prob} is formed backwards, which means that the direction of the protocol giving rise to such a distribution is in contrast to the thermodynamic scenario. While the mathematical origin of this discrepancy is clear, its deeper physical meaning remains elusive. As remarked earlier in the text, from the analogue fluctuation theorems in Eqs.~\eqref{Jarzynski} and \eqref{CohCrooks}, it can also be seen that the partition function of statistical mechanics is akin to the diagonal rank of states in coherence theory, as both quantifiers affect the availability of the given resource, i.e., work in thermodynamics and coherence in this paper.

The results in our study also have important implications for the resource theory of quantum coherence in its own right \cite{baumgratz2014quantifying,streltsov2017colloquium}.
In this respect, it is worth comparing explicitly the power of different assisted and unassisted scenarios for coherence distillation and state manipulations under incoherent operations.

In the standard unassisted scenario, as depicted in Figure~\ref{fig:circuit}{(a)}, the majorisation criteria recalled in Section~\ref{StateTransformation} can be equivalently formulated in terms of relative entropy: the state transformation $\ket{\psi}_A \rightarrow \ket{\phi}_A$ is possible by deterministic IO (or SIO) if and only if \cite{Du2015b,Du2017,winter2016operational,Chitambar2016,Chitambar2016b,Chitambar2017,Zhu2017}
\begin{equation}\label{zerolaw}
C_{\text{rel}}(\psi_{A}) \geq C_{\text{rel}}(\phi_{A})\,.
\end{equation}

If we consider instead the framework where the $d$-dimensional system $A$ is assisted by an exact catalyst $B$, whose state $\tau_B$ needs to be returned unchanged, then the state transformation $\ket{\psi}_A \otimes \ket{\tau}_B \rightarrow \ket{\phi}_A \otimes \ket{\tau}_B$ is possible by deterministic IO if and only if \cite{Bu2015b}
\begin{equation}
\frac{S_{\alpha}\big(\Delta(\psi_A)\big)-\ln d}{|\alpha|} > \frac{S_{\alpha}\big(\Delta(\phi_A)\big)-\ln d}{|\alpha|}\,, \quad \forall\,\alpha \in (-\infty,\infty)\,,
\end{equation}
where the quantities $S_{\alpha}(\rho) \coloneqq \text{sgn}(\alpha) \ln \big(\text{Tr} (\rho^\alpha)\big)/(1-\alpha)$ denote a family of R\'enyi entropies.

In the paradigm investigated in this paper, illustrated in Figure~\ref{fig:circuit}{(b)}, the system $A$ and the battery $B$ are still prepared in an initial product state $\ket{\psi}_A \otimes \ket{\lambda}_B$, but the coherence in the battery $B$ is allowed to change by an amount $w$ with probability $P(w)$, so that the battery plays the role of an approximate catalyst. In this case, we have shown that {\it any} transformation of the form (\ref{transformation}) can be implemented by IO or SIO on the system and the battery, provided the extracted coherence obeys the second law (\ref{first}),
\begin{align}\label{firstlaw}
	 C_{\text{rel}}(\psi_{A})-C_{\text{rel}}(\phi_{A}) \geq \langle w \rangle\,.
\end{align}
This is somehow comparable to the quantum thermodynamic setting recently investigated in \cite{mueller}, in which by allowing the buildup of arbitrarily small correlations between a system and a catalyst during a thermal operation, one finds that state transformations (for states diagonal in the energy eigenbasis) are specified by the second law expressed just in terms of the conventional Helmholtz free energy \cite{mueller}, rather than in terms of a whole family of R\'enyi entropies \cite{Brandao2015a}.

In the standard resource theory of coherence, it is currently understood that pure state transformations are reversible only in the asymptotic setting of many copies \cite{streltsov2017colloquium}. However, if a coherence battery is employed, then according to the coherence fluctuation theorem in Eq.~\eqref{unexpanded} a single copy transformation becomes reversible, when accompanied by a nontrivial fluctuation in coherence with probability $P(w)$. This means that, if there exists a coherence fluctuation $w=\ln q_i -\ln p_i $, then there exists a reverse process with equal and opposite coherence $w_{\text{rev}}=\ln p_i-\ln q_i$, where the forward protocol is nevertheless exponentially preferred according to the analogue Crooks relation in Eq.~(\ref{CohCrooks}).

Within the standard (unassisted) coherence resource theory, it is also known that in the asymptotic setting the distillable coherence under IO for any state is given by the relative entropy of coherence \cite{Winter2016}. However, comparing Eqs.~(\ref{zerolaw}) and (\ref{firstlaw}), we may conclude that  the second law of coherence demonstrates that, although on average the distillable coherence of a state is given by the relative entropy, in the battery assisted framework there is an exponentially suppressed regime in which more coherence can be extracted.

As a next step, it would be desirable to generalise our analysis to the non-asymptotic regime and explore the role of coherence fluctuation relations in the context of one-shot state transformations under different classes of incoherent operations, following recent work on one-shot coherence dilution and distillation \cite{oneshotdilution,oneshotdistillation}, and inspired by thermodynamic studies of one-shot dissipated work \cite{bobo,nicole}.

We have not considered the alternative assisted framework where the initial state $\rho_{AB}$ contains correlations between system $A$ and ancilla $B$, as that case would require the state of the system to be mixed. It is known that, in such a collaborative context, the asymptotic distillable coherence on  $A$ under local incoherent-quantum operations amounts to $S(\Delta(\rho_A))$, which yields a net gain over the unassisted case $C_{\text{rel}}(\rho_A)$ by a quantity equal to the reduced von Neumann entropy $S(\rho_A)$ \cite{Chitambar2015}. It may be worthy in the future to investigate coherence fluctuations in the battery, enhanced by initial correlations with the system. This would be especially interesting in view of the fact that the laws of thermodynamics in the presence of correlations (which might allow for seemingly paradoxical feats such as anomalous heat flow) have only recently begun to be understood in terms of physical processes \cite{Manab2017}.

It is hoped that, in the same way that the fluctuation theorems of statistical mechanics and thermodynamics have opened up a wide range of theoretic and experimental investigations, the fluctuation theorems of quantum coherence may inspire the discovery of new phenomena within coherence theory and applications, and beyond. It has already been shown in this work that fluctuating coherence allows one to break current limitations on reversibility and distillation in state transformations. The study of hybrid frameworks whereas a coherence battery may be employed to assist state transformations in different resource theories, such as athermality, entanglement, and more general manifestations of non-classicality, also deserves further investigation. This could complement recent studies of catalytic coherence for work extraction \cite{Aberg14,Korzekwa2015} and reveal new crossing points between the characterisation of coherence and quantum correlations \cite{Streltsov2015b,Zhu2017,MaQC} in quantum information theory.

\section*{Acknowledgements}

We warmly thank Bartosz Regula, Ludovico Lami, Alexander Streltsov, Luis A. Correa, and Tommaso Tufarelli for very fruitful discussions on the topic of this work and valuable feedback on earlier versions of the manuscript. We also acknowledge useful conversations with Alvaro Alhambra and Christopher Perry during QIP 2018. This work was supported by the European Research Council (ERC) under the Starting Grant GQCOP (Grant No.~637352) and the Foundational Questions Institute (fqxi.org) under the Physics of the Observer Programme (Grant No.~FQXi-RFP-1601).

\appendix

\section{Proof of necessary and sufficient conditions for B(S)IO state transformations} \label{suff}
Here we complete the proof of the Theorem  in Section~\ref{PureTrans}.

\subsection{Sufficient conditions\label{A1}}

Starting with the conditional probability distribution $P(i,w|j)$ satisfying Conditions 1--3 as discussed in Section~\ref{PureTrans}, let us define a sequence of initial and final states of the $d$-dimensional system $A$ and the battery $B$, indexed by an even $N$, as follows,
\begin{align}\label{initial}
|\Psi^{(N)}\rangle_{AB}=\sum_{i=1}^{d}\sum_{x=0}^{n}\sqrt{\sum_w\frac{p_{i,w}}{N+1}\beth_{x+f_w}({\cal S})}\ |i\rangle_A\otimes|c_x\rangle_B,
\end{align}
\begin{align}\label{finaln}
|\Phi^{(N)}\rangle_{AB}=\sum_{i=1}^{d}\sum_{x'=0}^{n}\sqrt{\sum_w\frac{q_{j}}{N+1}}\beth_{x'}({\cal S})\ |j\rangle_A\otimes|c_{x'}\rangle_B.
\end{align}
Here $p_{i,w} \coloneqq \sum_{j=1}^d P(i,w|j)q_j$ and the integer factor
\begin{equation}\label{factor}f_w \coloneqq \left\lfloor \frac{w}{\delta w}\right\rfloor
\end{equation}
measures the fluctuating coherence $w$ in units of $\delta w$, while the symbol $\beth_{k}(\mathcal{S})$ with ${\cal S} \coloneqq \left\{\frac{n-N}{2},...,\frac{n+N}{2}\right\}$ denotes an indicator function, i.e., it equals $1$ if the index $k \in \mathcal{S}$ and $0$ otherwise. Furthermore, we set $N \coloneqq n - 2f_{\max}$, with $f_{\max} \coloneqq \displaystyle \max_w \{|f_w|\}$.

We then need to construct a sequence of bistochastic matrices $G^{(N)}$ mapping the (nonzero) diagonal entries of $\Phi^{(N)}_{AB}$ to those of $\Psi^{(N)}_{AB}$ \cite{bhatia_1996}. The following adapts the proof reported in \cite{alhambra2017entanglement} for entanglement theory, to which the reader is referred for further details.

Let us begin by defining a new probability distribution
\begin{align}\label{BigProb}
P(i,x|j,x')\coloneqq \sum_w P(i,w|j)\delta_{x'-x,f_w},
\end{align}
with $x,x' \in (-\infty,\infty)$. The three conditions (\ref{Condition 1})--(\ref{Condition 3}) can then be rewritten as
\begin{align}
\sum_{i=1}^{d}\sum_{x=-\infty}^{\infty}P(i,x|j,x')=&1\,,\label{newconprob1}\\
\sum_{i=1}^{d}\sum_{x=-\infty}^{\infty}P(i,x|j,x')\Big{(}\frac{u}{u-1}\Big{)}^{x'-x}\le&1\,,\label{newconprob2}\\
\sum_{j=1}^{d}\sum_{x'=-\infty}^{\infty}P(i,x|j,x')\frac{q_j}{N+1}\beth_{x'}({\cal S})=
&\sum_w\frac{p_{i,w}}{N+1}\beth_{x+f_w}({\cal S})\,,\label{newconprob3}
\end{align}
where for the second equation we have used the fact that $f_w$ is the largest integer such that $f_w \ln\left(\frac{u}{u-1}\right)\leq w$, which follows from the definitions (\ref{deltaw}) and (\ref{factor}).

With this newly defined conditional probability $P(i,x|j,x')$, a sequence of sub-bistochastic matrices can now be constructed, whose rows and columns are labelled by the reference product basis of system and battery, say $\ket{i,z}_{AB}$ and $\ket{j,z'}_{AB}$, respectively. Here and in the following we employ the shorthand notation $z \in \chi_x$ to mean $\ket{z} \in \text{supp}(\chi_x)$. Then, for all $z\in \chi_x, \,z'\in \chi_{x'}$, where now the values of $x,x'$ are truncated as $x,x'\in \{0,...,n\}$,  let
\begin{align}\label{Rmatrix}
R^{(N)}(i,z|j,z') \coloneqq P(i,x|j,x')u^{-x}(u-1)^{x-n}\,.
\end{align}
Using Eqs.~\eqref{newconprob1}--\eqref{newconprob3}, we see that the matrices $R^{(N)}(i,z|j,z')$ are not bistochastic, but they already possess the desired property of mapping the entries of  $\Delta(\Phi^{(N)})$ to those of $\Delta(\Psi^{(N)})$,
\begin{align}
&\sum_{j=1}^{d}\sum_{x'=\frac{n-N}{2}-f_{\max}}^{\frac{n+N}{2}+f_{\max}}\sum_{z\in \chi_{x'}}R(i,z|j,z')\frac{q_j}{N+1}u^{-x'}(u-1)^{x'-n}\beth_{x'}({\cal S})\nonumber\\
=&\sum_{j=1}^{d}\sum_{x'=-\infty}^{\infty}P(i,x|j,x')\frac{q_j}{N+1}u^{-x}(u-x)^{x-n}\beth_{x'}({\cal S})\nonumber\\
=&\sum_w\frac{p_{i,w}}{N+1}u^{-x}(u-x)^{x-n}\beth_{x+f_w}({\cal S})\,. \label{subbi}
\end{align}

Finally, we can define a sequence of bistochastic matrices $G^{(N)}(i,z|j,z')$ based on $R^{(N)}(i,z|j,z')$, such that the action of the latter on the support of $\Delta(\Phi^{(N)})$ is left intact. This can be done as follows:
\begin{equation}\label{Gstring}
G^{(N)}(i,z|j,z')\coloneqq \left\{
                             \begin{array}{ll}
                               R^{(N)}(i,z|j,z')\,, & \hspace*{-3cm} \hbox{$\forall\ z' \in \chi_{x'}$ with $x'\in  \{\frac{n-N}{2},...,\frac{n+N}{2}\}$\,;} \\
                               \frac{1}{d\mu}\Big(1- \sum_{j=1}^{d}\sum_{x'=\frac{n-N}{2}}^{\frac{n+N}{2}}\sum_{z'\in \chi_{x'}}R^{(N)}(i,z|j,z')\Big)\,, & \hbox{\ \ \ \ otherwise\,,}
                             \end{array}
                           \right.
\end{equation}
where $\mu \coloneqq \sum_{x'=0}^{\frac{n-N}{2}}u^{x'}(u-1)^{n-x'}+\sum_{x'=\frac{n+N}{2}}^{n}u^{x'}(u-1)^{n-x'}$, so that $d \mu$ is the number of columns of $G^{(N)}$ not belonging to  the support of $\Delta(\Phi_N)$. It follows by construction that the matrices $G^{(N)}(i,z|j,z')$ are bistochastic,
\begin{equation}
\sum_{i=1}^d \sum_{x=0}^n \sum_{z \in \chi_x} G^{(N)}(i,z|j,z') = \sum_{j=1}^d \sum_{x'=0}^n \sum_{z' \in \chi_x'} G^{(N)}(i,z|j,z') = 1\,.
\end{equation}
 Moreover, by virtue of Eq.~(\ref{subbi}), these matrices map the diagonal coefficients of $\ket{\Phi^{(N)}}_{AB}$ to those of $\ket{\Psi^{(N)}}_{AB}$. This means that, according to \cite{bhatia_1996}, there exists a sequence of (S)IO protocols  $\Lambda^{(N)}_{AB}$ that implement the state transformations $\Phi^{(N)}_{AB}=\Lambda^{(N)}_{AB}(\Psi^{(N)}_{AB})$.

We are only left to verify that  $\lim_{N \rightarrow \infty} \ket{\Psi^{(N)}}_{AB}=\ket{\psi}_A \otimes \ket{\lambda}_B$ and $\lim_{N \rightarrow \infty} \ket{\Phi^{(N)}}_{AB}=\ket{\phi}_A \otimes \ket{\lambda}_B$. For the final states $\ket{\Phi^{(N)}}_{AB}$, which are uncorrelated across the system versus battery split for any finite $N$, this is manifestly true. For the initial states, $\ket{\Psi^{(N)}}_{AB}$, which are instead entangled across such a split, we need to analyse the large $N$ limit explicitly. It is straightforward to show \cite{alhambra2017entanglement} that the overlaps between the states $\ket{\Psi^{(N)}}_{AB}$ of Eq.~(\ref{initial}) and the product states
\begin{align}
|\widetilde{\Psi}^{(N)}\rangle_{AB}=\sum_{i=1}^{d}\sum_{x=0}^{n}\sqrt{\frac{p_i}{n+1}}|i\rangle_A\otimes|c_x\rangle_B,
\end{align}
which reproduce the correct initial state in the limit $N \rightarrow \infty$, satisfy
\begin{align}
\langle\widetilde{\Psi}^{(N)}|\Psi^{(N)}\rangle&=\sum_{i=1}^{d}\sum_{x=0}^{n}\sqrt{\frac{p_i}{n+1}\sum_w\frac{p_{i,w}}{N+1}\beth_{x+f_w}({\cal S})}\nonumber\\
&\ge \sum_{i=1}^d\sum_{x=\frac{n-N}{2}+f_{\max}}^{\frac{n+N}{2}-f_{\max}}\sqrt{\frac{p_i^2}{(n+1)(N+1)}}\nonumber\\
&\ge \frac{1}{1+\frac{2f_{\max}}{N+1}}  \overset{N \rightarrow \infty}\longrightarrow 1\,.
\end{align}
This concludes the proof of the sufficient conditions for the B(S)IO state transformation (\ref{transformation}).

\subsection{Necessary conditions\label{A2}}
Here we show that the conditional probability distribution $P(i,w|j)$ defined in Eq.~(\ref{con_prob}) satisfies the three conditions detailed in Eqs.~\eqref{Condition 1}, \eqref{Condition 2}, and \eqref{Condition 3}.

Before proceeding, let us note that all the (diagonal) battery states considered in the protocol of Section~\ref{PureTrans} live entirely in the subspace ${\cal B}$ introduced in Section~\ref{battery}. Accordingly, the supports of $\Delta(\Psi_{AB})$ and $\Delta(\Phi_{AB})$ are contained in the relevant subspace ${\cal V} \coloneqq {\cal D}({\cal H}_A) \otimes {\cal B}$ of the state space of system and battery, and the whole protocol described in Section~\ref{PureTrans} to construct $P(i,w|j)$ maps this subspace into itself. It follows that, as
$\Delta(\Psi_{AB}) \prec \Delta(\Phi_{AB})$ by hypothesis,  then one can find permutations $\Xi_m$ and probabilities $r_m$ such that the unital map $\EE_{AB}$ of the form (\ref{map}) satisfies (\ref{EEab}) and furthermore leaves invariant the projector $\mathds{1}_A \otimes \mathds{B}_B$ onto ${\cal V}$.

To prove Condition 1 (normalisation), we sum over the initial reference basis $i$ and the extracted coherence $w$,
\begin{align}
\sum_{i,w}P(i,w|j)&=\sum_{x'}\text{Tr}\big{[}(\mathds{1}_A \otimes \mathds{B}_{B})\mathcal{E}_{AB}(|j\rangle \langle j|_A \otimes \Pi^{x'}_{B}\Delta(\lambda_{B})\Pi^{x'}_{B})\big{]}\nonumber\\
&=\text{Tr}\big{[}|j\rangle \langle j|_A \otimes\Delta(\lambda_{B})\big{]}=1,
\end{align}
where we have used the fact that $\sum_w \Pi_{B}^{x'-\frac{w}{\delta w}}=\mathds{B}_{B}$ and that both the  map $\EE$ and the map resulting from the application of the projection operators $X\rightarrow \sum_{x'}\Pi^{x'}X\Pi^{x'}$ are trace preserving. This proves Eq.~(\ref{Condition 1}).

We continue with the proof of  Condition 2. Exploiting Eq.~(\ref{deltaw}), we can expand
\begin{align}\label{expand}
\Pi^{x'-\frac{w}{\delta w}}&=u^{x'-\frac{w}{\delta w}}\,(u-1)^{n-({x'-\frac{w}{\delta w}})}\chi_{x'-\frac{w}{\delta w}}\nonumber\\
&=\Big{\{}u^{x'}\,(u-1)^{n-x'}\Big{\}}u^{-\frac{w}{\delta w}}(u-1)^{\frac{w}{\delta w}}\chi_{x'-\frac{w}{\delta w}}\nonumber\\
&=\Big{\{}u^{x'}\,(u-1)^{n-x'}\Big{\}}\Big{(}\frac{u-1}{u}\Big{)}^{\frac{w}{\delta w}}\chi_{x'-\frac{w}{\delta w}}\nonumber\\
&=\Big{\{}u^{x'}\,(u-1)^{n-x'}\Big{\}}e^{-w}\chi_{x'-\frac{w}{\delta w}}\,.
\end{align}
Using the above relation we can write
\begin{align}
\sum_{w,j}P(i,w|j)e^w&=\sum_{w,x'}\alpha_{x'}\text{Tr}\big{[}(|i\rangle \langle i|_A\otimes {\chi_{x'-\frac{w}{\delta w}}}_B)\mathcal{E}_{AB}(\mathds{1}_A \otimes \Pi^{x'}_{B})\big{]}\ \nonumber \\
&\approx \sum_{w,x'}\alpha_{x'-\frac{w}{\delta w}}\text{Tr}\big{[}(|i\rangle \langle i|_A\otimes {\chi_{x'-\frac{w}{\delta w}}}_B)\mathcal{E}_{AB}(\mathds{1}_A \otimes \Pi^{x'}_{B})\big{]}\nonumber\\
&=\text{Tr}\big{[}(|i\rangle \langle i|_A\otimes \Delta(\lambda_{B}))\mathcal{E}_{AB}(\mathds{1}_A \otimes \mathds{B}_{B})\big{]} \nonumber \\
&=1\,,
\end{align}
where in the final line we have  used the property of the unital map $\mathcal{E}$ to preserve the projector on the diagonal support, while in the second line we have approximated $\alpha_{x'}\approx\alpha_{x'-\frac{w}{\delta w}}$. To verify the validity of the approximation, we invoke Eq.~(\ref{limit}), which yields
\begin{align}
\Big{|}\sum_{w,j}P(i,w|j)e^w-1\Big{|}&\leq  \sum_{w,x'} |\alpha_{x'} -\alpha_{x'-\frac{w}{\delta w}}| \nonumber \\
& \leq \sum_{w:|w| \leq w_{\max}} \sqrt{8\epsilon}\, \frac{|w|}{\delta w} \nonumber \\
& = \sqrt{8\epsilon}\, f_{\max}(f_{\max}+1)\,,
\end{align}
with $f_{\max}$ defined after Eq.~(\ref{factor}). This shows that Eq.~(\ref{Condition 2}) is fulfilled in the limit $\epsilon \rightarrow 0$.

We conclude by proving Condition 3, that expresses the correspondence between the marginal probabilities and the diagonal components of the states of the system.  Using the definition (\ref{con_prob2}) of the conditional probability, we can write
\begin{align}
\sum_{w,j}P(i,w|j)q_j&=\sum_{j,x'}q_j\alpha_{x'}\text{Tr}\big{[}(|i\rangle \langle i|_A\otimes \mathds{B}_{B})\mathcal{E}_{AB}(|j\rangle\langle j|_A \otimes {\chi_{x'}}_B)\big{]}\nonumber\\
&=\text{Tr}\big{[}(|i\rangle \langle i|_A\otimes \mathds{B}_{B})\mathcal{E}_{AB}(\Delta(\phi_{A} \otimes \lambda_{B}))\big{]}\nonumber\\
&\approx \text{Tr}\big{[}(|i\rangle \langle i|_A\otimes \mathds{B}_{B})\mathcal{E}_{AB}(\Delta(\Phi_{AB}))\big{]}\nonumber\\
&=\text{Tr}\big{[}(|i\rangle \langle i|_A\otimes \mathds{B}_{B})\Delta(\Psi_{AB})\big{]}\nonumber \\
&=p_i,
\end{align}
where we have used Eq.~(\ref{EEab}) and the approximation
\begin{equation}\label{deltapprox}
\Delta(\Phi_{AB})\approx\Delta(\phi_{A}\otimes \lambda_{B})=\bigg(\sum_jq_j|j\rangle \langle j |_B\bigg)\otimes\bigg(\sum_{x'}\alpha_{x'}{\chi_{x'}}_B\bigg)\,.
\end{equation}
To verify the validity of the approximation, we invoke Eq.~(\ref{close}), which yields
\begin{align}
\sum_i\bigg|\sum_{w,j}P(i,w|j)q_j-p_i\bigg|&=\sum_i\big|\text{Tr}\big{[}(|i\rangle \langle i|_A\otimes \mathds{B}_{B})\mathcal{E}_{AB}(\Delta(\phi_{A} \otimes \lambda_{B})-\Delta(\Phi_{AB}))\big{]}\big|\nonumber\\
&\leq \frac12 ||\mathcal{E}_{AB}(\Delta(\phi_{A} \otimes \lambda_{B}-\Phi_{AB}))||_1\nonumber\\
&\leq \frac12 ||\phi_{A} \otimes \lambda_{B}-\Phi_{AB}||_1\nonumber \\ &\leq\frac{\epsilon}{2},
\end{align}
where we have further used the contractivity of the trace distance under quantum channels and the fact that $\frac12 ||\rho-\sigma||_1= \displaystyle\max_{0\leq X \leq \mathds{1}} |\text{Tr} [X (\rho - \sigma)]|$ for two density matrices $\rho$ and $\sigma$. This shows that Eq.~(\ref{Condition 3}) is fulfilled as well in the limit $\epsilon \rightarrow 0$.

\section{Forward and reverse protocols for Crooks' coherence relation}\label{Crooks}

Here we construct and investigate both forward and reverse protocols for battery assisted (S)IO state transformations. The existence of a  (S)IO protocol for the state transformation $|\Psi\rangle_{AB} \rightarrow |\Phi\rangle_{AB}$, hereby referred to as {\it forward} protocol, is ensured by the majorisation relation $\Delta(\Psi_{AB}) \prec \Delta(\Phi_{AB})$, which we recall from Eq.~\eqref{trans} corresponds to
\begin{equation}\begin{aligned}
\Delta(\Psi_{AB}) = \sum_m r_m \Xi_m \Delta(\Phi_{AB}) \Xi^\dagger_m = \EE_{AB}(\Delta(\Phi_{AB})).
\end{aligned}\end{equation}
Let us now define a state $\ket{\Psi'}_{AB}$ with the same diagonal support as $\ket{\Phi}_{AB}$ and a state $\ket{\Phi'}_{AB}$ with the same diagonal support as $\ket{\Psi}_{AB}$, such that the existence of a (S)IO protocol for the state transformation $|\Psi'\rangle_{AB} \rightarrow |\Phi'\rangle_{AB}$, hereby referred to as {\it reverse} protocol, is ensured by the dual map,
\begin{equation}\begin{aligned}\label{eq:defined-state}
\Delta(\Psi'_{AB}) = \sum_m r_m \Xi_m^\dagger \Delta(\Phi'_{AB}) \Xi_m = \EE_{AB}^*(\Delta(\Phi'_{AB}))\,,
\end{aligned}\end{equation}
from which it is explicit that $\Delta(\Psi'_{AB}) \prec \Delta(\Phi'_{AB})$.

According to the B(S)IO framework, as we have seen in Section~\ref{PureTrans}, the forward protocol gives rise to a conditional probability distribution $P(i,w|j)$, which takes into account the fact that the coherence in the battery can fluctuate by an amount $w$ with probability $P(w)$.
Analogously, we will consider that in the reverse protocol the coherence in the battery is allowed to change by an amount $-w$ with probability $P^{\text{rev}}(-w)$.

Our aim is to compare the coherence distributions in the forward and reverse protocols. To do so, let us adopt the setting of \ref{A1}, and consider a sequence of forward (S)IO protocols implementing the transformations $|\Psi^{(N)}\rangle_{AB} \rightarrow  |\Phi^{(N)}\rangle_{AB}$, with initial and final states defined in Eqs.~(\ref{initial}) and (\ref{finaln}). In \ref{A1} we have introduced a sequence of bistochastic matrices $G^{(N)}$ mapping the (nonzero) diagonal coefficients of $\Phi^{(N)}_{AB}$ to those of $\Psi^{(N)}_{AB}$. We will use the fact that, by construction, these matrices satisfy
\begin{align}
G^{(N)}(i,z|j,z') = \text{Tr}[(|i\rangle\langle i|_{A} \otimes |z\rangle \langle z|_B) \EE_{AB}(|j\rangle\langle j|_A \otimes |z'\rangle \langle z'|_B)]\,.
\end{align}

Let us now define the transition matrix
\begin{equation}
Q(i,x|j,x')\coloneqq \text{Tr}[(|i\rangle\langle i|_A \otimes \Pi_B^x)\mathcal{E}_{AB}(|j\rangle\langle j|_A\otimes {\chi_{x'}}_B)]\,.\label{forward}
\end{equation}
From now on, we will assume for simplicity that $\frac{w}{\delta w}$ is an integer, i.e., $f_w \equiv \frac{w}{\delta w}$. Furthermore, we will impose $x' \in {\cal S}$, that is, we will limit the index $x'$ to span within the range in which the diagonal component of the battery has support. In this range, we have specifically
\begin{align}\label{defineQ}
Q(i,x|j,x')=&\sum_{z\in \chi_x,z'\in \chi_{x'}}u^{-x'}(u-1)^{x'-n}\text{Tr}[(|i\rangle\langle i|_A \otimes |z\rangle\langle z|_B)\mathcal{E}_{AB}(|j\rangle\langle j|_A \otimes |z'\rangle \langle z'|_B)]\nonumber\\
=&\sum_{z\in \chi_x, z'\in \chi_{x'}}G^{(N)}(i,z|j,z')\nonumber\\
=&\sum_{z\in \chi_x,z'\in \chi_{x'}}u^{-x'}(u-1)^{x'-n}P(i,x|j,x')\nonumber\\
=&P(i,x|j,x') \nonumber \\
=&\sum_w P(i,w|j)\delta_{x'-x,\frac{w}{\delta w}},
\end{align}
where in the last line we have used Eq.~(\ref{BigProb}). This means that, in the relevant range of $x'$, $Q(i,x|j,x')$ is directly related with the bistochastic matrix mapping the diagonal state coefficients in the forward protocol.

Inspired by Eq.~(\ref{eq:defined-state}), we can analogously define a transition matrix for the reverse protocol, given by
\begin{equation}
Q^{\text{rev}}(j,x'|i,x)\coloneqq \text{Tr}[(|j\rangle\langle j|_A\otimes \Pi_B^{x'})\mathcal{E}^{*}_{AB}(|i\rangle\langle i|_A\otimes {\chi_x}_B)]\,. \label{reverse}
\end{equation}
Noting that
\begin{align}\label{ratio}
Q^{\text{rev}}(j,x'|i,x)=&\frac{u^{x'}(u-1)^{n-x'}}{u^{x}(u-1)^{n-x}}Q(i,x|j,x')
\,,
\end{align}
this can be rewritten as
\begin{align}\label{Qrev}
Q^{\text{rev}}(j,x'|i,x)=&\frac{u^{x'}(u-1)^{n-x'}}{u^{x}(u-1)^{n-x}}\sum_w P(i,w|j)\delta_{x'-x,\frac{w}{\delta w}}\nonumber\\
=&\frac{u^{x'}(u-1)^{n-x'}}{u^{x}(u-1)^{n-x}}\sum_w \delta_{x'-x,\frac{w}{\delta w}}\sum_{x''}\alpha_{x''}\text{Tr}[(|i\rangle\langle i |_A\otimes \Pi_B^{x''-\frac{w}{\delta w}})\mathcal{E}_{AB}(|j\rangle\langle j|_A\otimes {\chi_{x''}}_B)]\nonumber \\
=&\sum_w \delta_{x'-x,\frac{w}{\delta w}}\sum_{x''}\Big{(}\frac{u}{u-1}\Big{)}^{\frac{w}{\delta w}}\alpha_{x''}\text{Tr}[(|i\rangle\langle i |_A\otimes \Pi_B^{x''-\frac{w}{\delta w}})\mathcal{E}_{AB}(|j\rangle\langle j|_A\otimes {\chi_{x''}}_B)]\nonumber\\
=&\sum_w \delta_{x'-x,\frac{w}{\delta w}}\sum_{x''}\alpha_{x''}\text{Tr}[(|j\rangle\langle j |_A\otimes \Pi_B^{x''})\mathcal{E}^{*}_{AB}(|i\rangle\langle i|_A\otimes {\chi_{x''-\frac{w}{\delta w}}}_B)].
\end{align}
This is now approximated to an ideal battery, which gives
\begin{align}
Q^{\text{rev}}(j,x'|i,x)\approx&\sum_w \delta_{x'-x,\frac{w}{\delta w}}\sum_{x''}\alpha_{x''+\frac{w}{\delta w}}\text{Tr}[(|j\rangle\langle j |_A\otimes \Pi_B^{x''+\frac{w}{\delta w}})\mathcal{E}^{*}_{AB}(|i\rangle\langle i|_A\otimes {\chi_{x''}}_B)]\nonumber\\
\approx&\sum_w \delta_{x'-x,\frac{w}{\delta w}}\sum_{x''}\alpha_{x''}\text{Tr}[(|j\rangle\langle j |_A\otimes \Pi_B^{x''+\frac{w}{\delta w}})\mathcal{E}^{*}_{AB}(|i\rangle\langle i|_A\otimes {\chi_{x''}}_B)]\nonumber\\
=&\sum_w P^{\text{rev}}(j,-w|i)\delta_{x'-x,\frac{w}{\delta w}}\,,
\end{align}
where we have defined the conditional probability distribution for the reverse protocol as
\begin{equation}\label{con_prob2_rev}
P^{\text{rev}}(j,-w|i) \coloneqq \sum_{x''}\alpha_{x''}\text{Tr}[(j\rangle\langle j |_A\otimes \Pi_B^{x''+\frac{w}{\delta w}})\mathcal{E}^{*}_{AB}(|i\rangle\langle i|_A\otimes {\chi_{x''}}_B)]\,,
\end{equation}
The  distribution (\ref{con_prob2_rev}) is the counterpart to the distribution $P(i,w|j)$ given in  (\ref{con_prob2}) for the forward protocol, and it satisfies analogous conditions to those proved in \ref{A2}, i.e.,
\begin{align}
\sum_{j,w}P^{\text{rev}}(j,-w|i)=&1\,,\\
\sum_{i,w}P^{\text{rev}}(j,-w|i) e^{-w}=&1\,,\\
\sum_{i,w}P^{\text{rev}}(j,-w|i)q'_i=&p'_j\,,
\end{align}
where $p'_i$ and $q'_i$ respectively denote the diagonal elements of the initial and final states of the system $A$ in the reverse protocol.

Therefore, given a sequence of forward (S)IO protocols with transition matrix $Q(i,x|j,x')$ mapping the diagonal coefficients of final states $|{\Phi}^{(N)}\rangle_{AB}$ to those of initial states $|{\Psi}^{(N)}\rangle_{AB}$, there exists a sequence of reverse (S)IO protocols with transition matrix $Q^{\text{rev}}(j,x'|i,x)$ mapping the diagonal coefficients of final states $|{\Phi'}^{(N)}\rangle_{AB}$ to those of initial states $|{\Psi'}^{(N)}\rangle_{AB}$.
The initial and final states for the reverse protocol can then be chosen, in analogy to the analysis of \ref{A1}, as
\begin{align}
|{\Psi'}^{(N)}\rangle_{AB} &= \sum_{j=1}^d \sum_{x'=0}^n\sqrt{\frac{p_{j,-w}'}{N'+1}\beth_{x'-f_w}({\cal S}')}\  |j\rangle_A\otimes|c_{x'}\rangle_B\,, \label{initialrev} \\
|{\Phi'}^{(N)}\rangle_{AB} &= \sum_{i=1}^d\sum_{x=0}^n\sqrt{\frac{q_i'}{N'+1}}\beth_{x}({\cal S}')\ |i\rangle_A \otimes|c_x\rangle_B\,, \label{finalnrev}
\end{align}
where $p'_{j,-w} \coloneqq \sum_{i=1}^d P^{\text{rev}}(j,-w|i) q'_i$,  ${\cal S}' \coloneqq \left\{\frac{n-N'}{2},...,\frac{n+N'}{2}\right\}$, and $N' \coloneqq N-2 f_{\max}$, thus ensuring that $x' \in {\cal S}$. It is easy then to verify that Eq.~(\ref{eq:defined-state}) is satisfied for the reverse protocol in the limit $N \rightarrow \infty$, adapting the derivation of \ref{A1}.

The coherence analogue of Crooks' theorem can now be obtained. In Crooks' theorem both the initial and final states are thermal, here to derive its equivalent the final states of both the forward and reverse protocols will be taken to be maximally coherent (with diagonal rank $d'$ and $d$ respectively), which sets $q_j=\frac{1}{d'}$ and $q_{i}'=\frac{1}{d}$.

The forward and reverse coherence distributions are therefore
\begin{align}
P(w)=&\sum_{i,j}P(i,w|j)\frac{1}{d'}\\
P^{\text{rev}}(-w)=&\sum_{i,j}P^{\text{rev}}(j,-w|i)\frac{1}{d}.
\end{align}
Now recalling the definition (\ref{proj}) with $x=x'-\frac{w}{\delta w}$, and expanding as in (\ref{expand}), we can rewrite Eq.~(\ref{ratio}) simply as
\begin{align}\label{ratiow}
Q^{\text{rev}}(j,x'|i,x)= e^w Q(i,x|j,x')\,,
\end{align}
from which it is immediate to see that the above two distributions $P(w)$ and $P^{\text{rev}}(-w)$ obey Crooks' analogue relation given in Eq.~(\ref{CohCrooks}).



\section*{References}


\bibliographystyle{iopart-num}
\bibliography{Reference_Index}

\providecommand{\newblock}{}
\begin{thebibliography}{10}
\expandafter\ifx\csname url\endcsname\relax
  \def\url#1{{\tt #1}}\fi
\expandafter\ifx\csname urlprefix\endcsname\relax\def\urlprefix{URL }\fi
\providecommand{\eprint}[2][]{\url{#2}}

\bibitem{streltsov2017colloquium}
Streltsov A, Adesso G and Plenio M~B 2017 {\em Rev. Mod. Phys.\/} {\bf 89}
  041003

\bibitem{aaberg2006quantifying}
\AA{}berg J 2006 (\textit{Preprint} \eprint{quant-ph/0612146})

\bibitem{baumgratz2014quantifying}
Baumgratz T, Cramer M and Plenio M 2014 {\em Phys. Rev. Lett.\/} {\bf 113}
  140401

\bibitem{winter2016operational}
Winter A and Yang D 2016 {\em Phys. Rev. Lett.\/} {\bf 116} 120404

\bibitem{Streltsov2015b}
Streltsov A, Singh U, Dhar H~S, Bera M~N and Adesso G 2015 {\em Phys. Rev.
  Lett.\/} {\bf 115}(2) 020403

\bibitem{Napoli2016}
Napoli C, Bromley T~R, Cianciaruso M, Piani M, Johnston N and Adesso G 2016
  {\em Phys. Rev. Lett.\/} {\bf 116}(15) 150502
  \urlprefix\url{http://link.aps.org/doi/10.1103/PhysRevLett.116.150502}

\bibitem{Chitambar2016}
Chitambar E and Gour G 2016 {\em Phys. Rev. Lett.\/} {\bf 117}(3) 030401

\bibitem{Chitambar2016b}
Chitambar E and Gour G 2016 {\em Phys. Rev. A\/} {\bf 94}(5) 052336
  \urlprefix\url{http://link.aps.org/doi/10.1103/PhysRevA.94.052336}

\bibitem{Chitambar2017}
Chitambar E and Gour G 2017 {\em Phys. Rev. A\/} {\bf 95}(1) 019902
  \urlprefix\url{http://link.aps.org/doi/10.1103/PhysRevA.95.019902}

\bibitem{Zhu2017}
Zhu H, Ma Z, Cao Z, Fei S~M and Vedral V 2017 {\em Phys. Rev. A\/} {\bf 96}(3)
  032316 \urlprefix\url{https://link.aps.org/doi/10.1103/PhysRevA.96.032316}

\bibitem{Wang:17}
Wang Y~T, Tang J~S, Wei Z~Y, Yu S, Ke Z~J, Xu X~Y, Li C~F and Guo G~C 2017 {\em
  Phys. Rev. Lett.\/} {\bf 118}(2) 020403
  \urlprefix\url{https://link.aps.org/doi/10.1103/PhysRevLett.118.020403}

\bibitem{Wu:17}
Wu K~D, Hou Z, Zhong H~S, Yuan Y, Xiang G~Y, Li C~F and Guo G~C 2017 {\em
  Optica\/} {\bf 4} 454

\bibitem{carnot1872reflexions}
Carnot S 1872 R{\'e}flexions sur la puissance motrice du feu et sur les
  machines propres {\`a} d{\'e}velopper cette puissance (Annales scientifiques
  de l'Ecole normale)

\bibitem{Kosloff}
Kosloff R 2013 {\em Entropy\/} {\bf 15} 2100 ISSN 1099-4300
  \urlprefix\url{http://www.mdpi.com/1099-4300/15/6/2100}

\bibitem{Goold2016}
Goold J, Huber M, Riera A, del Rio L and Skrzypczyk P 2016 {\em J. Phys. A\/}
  {\bf 49} 143001

\bibitem{jarzynski1997nonequilibrium}
Jarzynski C 1997 {\em Phys. Rev. Lett.\/} {\bf 78} 2690

\bibitem{crooks1999entropy}
Crooks G~E 1999 {\em Phys. Rev. E\/} {\bf 60} 2721

\bibitem{alhambra2016fluctuating}
Alhambra {\'A}~M, Masanes L, Oppenheim J and Perry C 2016 {\em Phys. Rev. X\/}
  {\bf 6} 041017

\bibitem{alhambra2017entanglement}
Alhambra {\'A}~M, Masanes L, Oppenheim J and Perry C 2017 {\em arXiv preprint
  arXiv:1709.06139\/}

\bibitem{horodecki2013fundamental}
Horodecki M and Oppenheim J 2013 {\em Nat. Commun.\/} {\bf 4} 2059
  \urlprefix\url{http://dx.doi.org/10.1038/ncomms3059}

\bibitem{Gour2013}
Gour G, M{\"u}ller M~P, Narasimhachar V, Spekkens R~W and Halpern N~Y 2015 {\em
  Phys. Rep.\/} {\bf 583} 1 -- 58 ISSN 0370-1573

\bibitem{nielsen1999conditions}
Nielsen M~A 1999 {\em Phys. Rev. Lett.\/} {\bf 83} 436

\bibitem{yadin2016quantum}
Yadin B, Ma J, Girolami D, Gu M and Vedral V 2016 {\em Phys. Rev. X\/} {\bf 6}
  041028

\bibitem{Herbut2005}
Herbut F 2005 {\em J. Phys. A\/} {\bf 38} 2959

\bibitem{Vaccaro2008}
Vaccaro J~A, Anselmi F, Wiseman H~M and Jacobs K 2008 {\em Phys. Rev. A\/} {\bf
  77}(3) 032114
  \urlprefix\url{http://link.aps.org/doi/10.1103/PhysRevA.77.032114}

\bibitem{Gour2009}
Gour G, Marvian I and Spekkens R~W 2009 {\em Phys. Rev. A\/} {\bf 80}(1) 012307

\bibitem{Du2015b}
Du S, Bai Z and Guo Y 2015 {\em Phys. Rev. A\/} {\bf 91}(5) 052120

\bibitem{Du2017}
Du S, Bai Z and Guo Y 2017 {\em Phys. Rev. A\/} {\bf 95}(2) 029901
  \urlprefix\url{http://link.aps.org/doi/10.1103/PhysRevA.95.029901}

\bibitem{bhatia_1996}
Bhatia R 1996 {\em Matrix {{Analysis}}\/} ({Springer}) ISBN 978-0-387-94846-1

\bibitem{nielsen_2001}
Nielsen M~A and Vidal G 2001 {\em Quantum Info Comput\/} {\bf 1} 76--93 ISSN
  1533-7146 \urlprefix\url{http://dl.acm.org/citation.cfm?id=2011326.2011331}

\bibitem{masanes2017general}
Masanes L and Oppenheim J 2017 {\em Nat. Commun.\/} {\bf 8} 14538

\bibitem{Brandao2015a}
Brand{\~a}o F~G~S~L, Horodecki M, Ng N~H~Y, Oppenheim J and Wehner S 2015 {\em
  Proc. Natl. Acad. Sci. U.S.A.\/} {\bf 112} 3275--3279

\bibitem{ABC}
Adesso G, Bromley T~R and Cianciaruso M 2016 {\em J. Phys. A\/} {\bf 49} 473001

\bibitem{nonextensive}
Abe S and Rajagopal A~K 2003 {\em Phys. Rev. Lett.\/} {\bf 91}(12) 120601

\bibitem{Bu2015b}
Bu K, Singh U and Wu J 2016 {\em Phys. Rev. A\/} {\bf 93}(4) 042326
  \urlprefix\url{http://link.aps.org/doi/10.1103/PhysRevA.93.042326}

\bibitem{mueller}
M\"uller M~P 2017 {\em arXiv preprint arXiv:1707.03451\/}

\bibitem{Winter2016}
Winter A and Yang D 2016 {\em Phys. Rev. Lett.\/} {\bf 116} 120404

\bibitem{oneshotdilution}
Zhao Q, Liu Y, Yuan X, Chitambar E and Ma X 2018 {\em Phys. Rev. Lett.\/} {\bf
  120}(7) 070403

\bibitem{oneshotdistillation}
Regula B, Fang K, Wang X and Adesso G 2017 {\em arXiv preprint
  arXiv:1711.10512\/}

\bibitem{bobo}
Wei B~B and Plenio M~B 2017 {\em New J. Phys.\/} {\bf 19} 023002

\bibitem{nicole}
Halpern N~Y, Garner A~J~P, Dahlsten O~C~O and Vedral V 2015 {\em arXiv preprint
  arXiv:1505.06217\/}

\bibitem{Chitambar2015}
Chitambar E, Streltsov A, Rana S, Bera M~N, Adesso G and Lewenstein M 2016 {\em
  Phys. Rev. Lett.\/} {\bf 116}(7) 070402
  \urlprefix\url{http://link.aps.org/doi/10.1103/PhysRevLett.116.070402}

\bibitem{Manab2017}
Bera M~N, Riera A, Lewenstein M and Winter A 2017 {\em Nat. Commun.\/} {\bf 8}
  2180

\bibitem{Aberg14}
\AA{}berg J 2014 {\em Phys. Rev. Lett.\/} {\bf 113}(15) 150402

\bibitem{Korzekwa2015}
Korzekwa K, Lostaglio M, Oppenheim J and Jennings D 2016 {\em New J. Phys.\/}
  {\bf 18} 023045

\bibitem{MaQC}
Ma J, Yadin B, Girolami D, Vedral V and Gu M 2016 {\em Phys. Rev. Lett.\/} {\bf
  116}(16) 160407

\end{thebibliography}


\end{document}